\documentclass{ws-mpla}

\begin{document}

\def\dsp{\displaystyle}
\def\oneh{{\textstyle {\frac{1}{2}}}}
\newcommand{\tvec}[1]{\mbox{\boldmath{$#1$}}}
\newcommand{\svec}[1]{\mbox{\boldmath{$\scriptstyle #1$}}}
\newcommand{\nn}{\nonumber}
\newcommand{\be}{\begin{equation}}
\newcommand{\ee}{\end{equation}}
\newcommand{\ba}{\begin{eqnarray}}
\newcommand{\ea}{\end{eqnarray}}
\newcommand{\la}{\langle}
\newcommand{\ra}{\rangle}
\markboth{B.~Pasquini, S.~Boffi, P.~Schweitzer}
{The spin structure of the nucleon in light-cone quark models}

\catchline{}{}{}{}{}

\title{The spin structure of the nucleon in light-cone quark models}

\author{B. PASQUINI\footnote{Barbara.Pasquini@pv.infn.it}, 
S. BOFFI\footnote{Sigfrido.Boffi@pv.infn.it}}
\address{Dipartimento di Fisica Nucleare e Teorica, 
Universit\`a degli Studi di Pavia\\
Via Bassi 6, I-27110 Pavia, Italy, and\\Istituto Nazionale di Fisica Nucleare, 
Sezione di Pavia, Pavia, Italy
}
\author{P. SCHWEITZER
\footnote{peter.schweitzer@phys.uconn.edu}}

\address{Department of Physics, University of Connecticut, \\  
2152 Hillside Road, Storrs, CT  06269-3046, U.S.A.
}

\maketitle

\pub{Received (09 October 2009)}{}

\begin{abstract}
The quark spin densities related to generalized parton distributions in 
impact-parameter space and to transverse-momentum dependent parton distributions
 are reviewed within a light-cone quark model, with focus on the role
of the different spin-spin and spin-orbit correlations of quarks.
Results for azimuthal spin asymmetries in semi-inclusive deep-inelastic scattering due to T-even 
transverse-momentum dependent parton distributions are also discussed.

\keywords{parton correlation functions, light-cone quantization, semi-inclusive deep inelastic scattering}
\end{abstract}

\ccode{PACS Nos.: 12.39.Ki, 13.85.Ni, 13.60.-r   }

\section{Introduction}
One of the most challenging tasks for unravelling the partonic structure 
of hadrons is mapping the distribution of the spin of the proton 
onto its constituents.
To this aim, generalized parton distributions (GPDs)\cite{Mueller:1998fv}$^-$\cite{Boffi:2007yc} 
and transverse-momentum dependent
parton distributions (TMDs)\cite{Collins:1981uw,Mulders:1995dh}
 have proved to be among the most useful tools.
GPDs provide a new method of spatial imaging of the nucleon\cite{Burkardt:2005td,Diehl:2005jf}, through the 
definition of impact-parameter dependent spin densities  which reveal the 
correlations between the quark distributions in transverse-coordinate space 
and longitudinal momentum for different quark and target polarizations.
On the other hand, TMDs contain novel and direct three-dimensional information
about the strength of different spin-spin and spin-orbit correlations
in the momentum space.
Although GPDs and TMDs can be seen as two different limiting cases
of the same generalized parton-correlation functions, no-model independent 
relations between the two classes of objects has been obtained so 
far\cite{Meissner:2009ww,Meissner:2007rx}.
\newline
\noindent
A convenient way to make explicit which kind of information on hadron 
structure is contained in these quantities is the representation in terms 
of overlap of light-cone wave functions (LCWFs) which are the probability 
amplitudes to find a given $N$-parton configuration in the Fock-space 
expansion of the hadron state\cite{Brodsky:1997de}.
In the following, we will confine our analysis to the three-quark sector, 
by truncating the light-cone expansion of the nucleon state to the minimal
Fock-space configuration.
The three-quark component of the nucleon has been studied 
extensively in the literature\cite{Chernyak:1983ej}$^-$\cite{Pasquini:2009ki} 
in terms of quark 
distribution amplitudes defined as hadron-to-vacuum 
transition matrix elements of non-local gauge-invariant light-cone operators.
Unlike these works,
the authors of Refs.\cite{Burkardt:2002uc,Ji:2002xn} considered the 
wave-function amplitudes keeping full transverse-momentum dependence of 
partons and proposed a systematic way to enumerate independent amplitudes of 
a LCWF which parametrize the different orbital angular momentum components
of the nucleon state.
In particular, the three-quark LCWF involves 
six independent amplitudes corresponding to different combinations of quark orbital angular momentum and helicity.
With such amplitudes one can obtain a model-independent 
representation for the quark contribution to TMDs and GPDs which
emphasizes the role of the different orbital angular momentum components.
One could then choose a phenomenological approach 
parametrizing the light-cone amplitudes and fitting observables related to TMDs and GPDs  to data.
Here we will adopt 
a light-cone constituent quark model (CQM) which has been successfully 
applied in the calculation of the electroweak properties of the nucleon\cite{Pasquini:2007iz}.
As outlined in Ref.\cite{Boffi:2002yy},
the starting point is the three-quark wave function obtained as solution
of the  Schr\"odinger-like eigenvalue equation in the instant-form dynamics.
The corresponding solution in light-cone dynamics is obtained through the
unitary  transformation represented by product of Melosh rotations acting
on the spin of the individual quarks.
In particular, the instant-form wave function is constructed as a product of a momentum wave function which is spherically symmetric and invariant under permutations, and a spin-isospin wave function which is uniquely determined by SU(6)
symmetry requirements.
By applying the Melosh rotations, the Pauli spinors of the quarks in the
 nucleon rest frame are converted to light-cone spinors.
The relativistic spin effects are evident in the presence
of spin-flip terms in the Melosh rotations which generate non-zero orbital angular momentum components
and non-trivial correlations between spin and transverse momentum of the 
quarks. On the other hand,
the momentum-dependent wave function keeps the original functional form, with 
instant-form coordinates rewritten in terms of light-cone coordinates.
The explicit expressions of the light-cone amplitudes within this CQM 
 can be found in Ref.\cite{Pasquini:2008ax}, while the corresponding results for GPDs  in impact-parameter space and for TMDs will be discussed in 
sect. 2 and 3, respectively.
Finally, in sect. 4 we will show predictions  for
single spin asymmetries in semi-inclusive deep inelastic  scattering (SIDIS)
due to T-even TMDs.
\section{Spin densities in the impact parameter space}

In light cone gauge $A^+=0$, 
 GPDs are obtained from the same quark correlation function
entering the definition of the ordinary parton distributions, but now evaluated  between hadron  with 
different momentum  in the initial and final state.
They depend on the average longitudinal-momentum fraction $x$, the skewness parameter $\xi$ describing the longitudinal change of the nucleon momentum, 
and the momentum transfer $t=\Delta^2$.
When $\xi=0$ and $x>0$, by  a two-dimensional Fourier transform to 
impact-parameter space GPDs can be interpreted as densities of quarks with 
longitudinal momentum fraction $x$ and transverse location $\tvec b$ with 
respect to the nucleon center of momentum.
Depending on the polarization of both the active quark and the parent nucleon, one defines\cite{Burkardt:2005td,Diehl:2005jf} three-dimensional 
densities $\rho(x,{\tvec b}, \lambda,(\Lambda,\tvec{S}_T))$ and  $\rho(x,{\tvec b},{\tvec s}_T,{\tvec S}_T)$ representing the probability  to find a quark with longitudinal
 momentum fraction $x$ and transverse position $\tvec b$ either with 
light-cone helicity $\lambda$ ($=\pm 1$) 
or transverse spin $\tvec s_T$
in the nucleon with longitudinal
 polarization $\Lambda$ ($=\pm 1$) or transverse  spin $\tvec{S}_T$. 
They read
\begin{eqnarray}
\rho(x,{\tvec b}, \lambda,(\Lambda,\tvec{S}_T))) &=&  \oneh \left[ H(x,{b}^2) 
  + b^j\varepsilon^{ji} S^i_T  \frac{1}{M}\, 
       E'(x,{b}^2)
  + \lambda \Lambda \tilde{H}(x,{b}^2) \,\right] ,
 \label{eq:long}\\
\rho(x,{\tvec b},{\tvec s}_T,{\tvec S}_T) 
&= &{}\dsp \oneh\left[ H(x,{b}^2)  + s^i_TS^i_T\left( H_T(x,{b}^2)  -\frac{1}{4M^2} \Delta_b \tilde H_T(x,{b}^2) \right) \right.
\nonumber\\
& &\quad{}\dsp + \frac{b^j\varepsilon^{ji}}{M}\left(
S^i_TE'(x,{b}^2)  + s^i_T\left[ E'_T(x,{b}^2)  + 2 \tilde H'_T(x,{b}^2) \right]\right)
\nonumber\\
& &\quad\left.{}\dsp+ s^i_T(2b^ib^j - b^2\delta_{ij}) S^j_T\frac{1}{M^2} \tilde H''_T(x,{b}^2) \right],
 \label{eq:transv}
 \end{eqnarray}
where the derivatives are defined
$
f' = \frac{\partial}{\partial b^2}\, f $, and
$
\Delta_b f
= 4\, \frac{\partial}{\partial b^2}
    \Big( b^2 \frac{\partial}{\partial b^2} \Big) f $.
In Eqs.~(\ref{eq:long})-(\ref{eq:transv}) enter the Fourier transforms
of the GPDs for unpolarized quarks ($H$ and $E$), for longitudinally polarized quarks ($\tilde H$ and $\tilde E$) and transversely polarized quarks
 ($H_T$, $E_T$, $\tilde H_T$, and $\tilde E_T$). 
\newline
\noindent
In Eq.~(\ref{eq:long}) the first term with $H$ describes the density of 
unpolarized quarks in the unpolarized proton. The term with $E'$ introduces
 a sideways shift in such a density when the proton is transversely polarized, 
and the term with $\tilde H$ reflects the difference in the density of quarks 
with helicity equal or opposite to the proton helicity. 
\newline
\noindent
In the three lines of Eq.~(\ref{eq:transv}) one may distinguish the three
 contributions corresponding to monopole, dipole and quadrupole structures. 
The unpolarized quark density $\oneh H$ in the monopole structure is modified
 by the chiral-odd terms with $H_T$ and $\Delta_b \tilde H_T$ when both the 
quark and the proton are transversely polarized. Responsible for the dipole 
structure is either the same chiral-even contribution with $E'$ from the 
transversely polarized proton appearing in the  spin 
distribution~(\ref{eq:long}) or the chiral-odd contribution with 
$E'_T+2\tilde H'_T$ from the transversely polarized quarks or both. 
The quadrupole term with $\tilde H''_T$ is present only when both quark
 and proton are transversely polarized.
In terms of overlap of light-cone amplitudes, the monopole distributions
are associated to GPDs which are diagonal in the orbital angular momentum space, while the dipole distributions 
describe spin flip either of the nucleon or of the quark, and accordingly are given
by overlap of light-cone amplitudes which differ by one unit of orbital angular momentum in the initial and final state.
In the case of the chiral-odd GPD $\tilde H_T$ the nucleon helicity flips in the direction opposite
to the quark helicity, with a mismatch of orbital angular momentum of two units between the initial and final state.

\begin{figure}[t]
\centerline{\psfig{file=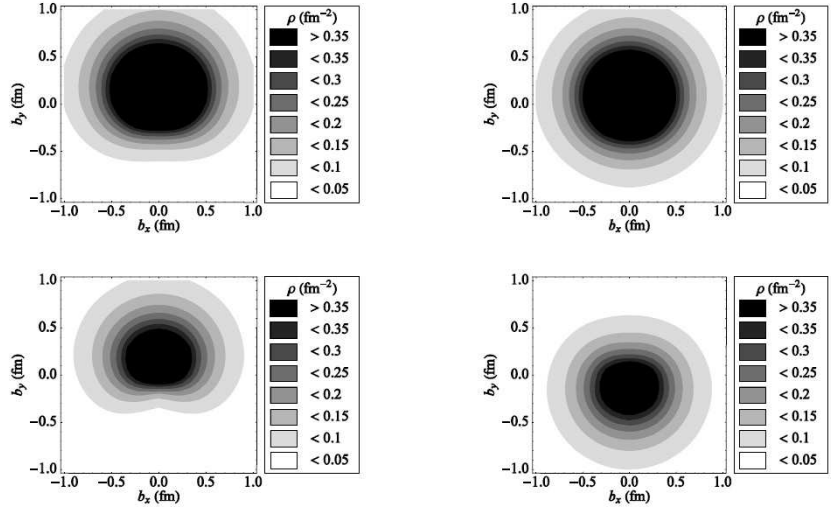,width=12.5 cm}}
\vspace*{8pt}
\caption{The spin-densities
for (transversely) $\hat x$-polarized quarks in an unpolarized proton
(left panels) and for unpolarized quarks in a  (transversely) $\hat x$-polarized proton.
The upper (lower) row corresponds to the results for up (down) quarks.
\protect\label{fig1}}
\end{figure}

In the following, 
we show some examples of  spin densities using the model predictions for the 
GPDs from Refs.\cite{Boffi:2002yy,Boffi:2003yj,Pasquini:2005dk}.
In the case of transversely polarized quarks in an unpolarized proton 
the dipole 
contribution introduces a large distortion perpendicular to both the quark 
spin and the momentum of the proton, as shown in the left column of
Fig.~\ref{fig1}. This effect has been 
related\cite{Burkardt:2005hp} to a non-vanishing Boer-Mulders
 function\cite{Boer:1997nt}  $h_1^\perp$ which describes the correlation between 
intrinsic transverse momentum and transverse spin of quarks. 
Such a distortion reflects the large value of  the anomalous tensor magnetic
 moment $\kappa_T$ for both flavors. Here, $\kappa^u_T=3.98$ 
and $\kappa^d_T=2.60$, to be compared with the values $\kappa^u_T\approx 3.0$ 
and $\kappa^d_T\approx 1.9$ of Ref.\cite{Gockeler:2006zu} due to a positive 
combination $E_T+2\tilde H_T$. Since $\kappa_T\sim - h_1^\perp$, the present 
results confirm the conjecture that $h_1^\perp$ is large and negative both 
for up and down quarks\cite{Burkardt:2005hp}.
As also noticed in Refs.\cite{Burkardt:2005hp,Gockeler:2006zu} the large anomalous 
magnetic moments $\kappa^{u,d}$ are responsible for the dipole distortion 
produced in the case of unpolarized quarks in transversely polarized
 nucleons (right column of Fig.~\ref{fig1}). 
With the present model, $\kappa^u=1.86$ and $\kappa^d=-1.57$, 
to be compared with the values $\kappa^u=1.673$ and $\kappa^d=-2.033$ 
derived from data. This effect can serve as a dynamical explanation of
 a non-vanishing Sivers function\cite{Sivers:1989cc} $f_{1T}^\perp$ 
which measures the correlation between the intrinsic quark transverse momentum 
and the transverse nucleon spin. The present results, with the opposite 
shift of up and down quark spin distributions imply an opposite sign of 
$f_{1T}^\perp$ for up and down quarks\cite{Burkardt:2005hp} as
 confirmed by the recent observation of the HERMES 
collaboration\cite{Airapetian:2004tw}.
The results in Fig.~\ref{fig1} are also in qualitative agreement with those obtained in lattice calculations\cite{Gockeler:2006zu}.

Finally,
we refer to\cite{Boffi:2007yc,Pasquini:2007xz}
 for the light-cone CQM results
of the densities with
more complex spin-configurations
with transverse  polarization of both the quark as well
as the proton.

\section{Transverse-momentum dependent distributions}
The eight leading-twist TMDs,
$f_1,$ $f_{1T}^{\perp },$ $g_1,$  $g_{1T},$ $g_{1L}^{\perp},$
$h_1,$  $h_{1T}^{\perp},$ $h_{1L}^{\perp}$, and 
$h_{1}^{\perp},$ are a natural extension of standard parton distribution from one to three dimensions in momentum space, being defined in terms of the 
same quark correlation functions but without integration over the transverse 
momentum.
 Among them,
 the Boer-Mulders $h_1^\perp$\cite{Boer:1997nt} and the Sivers $f_{1T}^\perp$\cite{Sivers:1989cc} functions are T-odd, 
i.e. they change sign under ``naive time reversal'', 
which is defined as usual time reversal, but without interchange of initial 
and final states.
Since non-vanishing T-odd TMDs require gauge boson degrees of freedom which are not taken into account in our light-cone quark model, 
our model results will be discussed only for the T-even TMDs.
\newline
\noindent
Projecting the correlator for quarks of definite longitudinal or transverse polarizations, one obtains the following 
spin densities in the momentum space
\begin{eqnarray}
  \label{piet-distr1}
\tilde\rho(x,\tvec{k}_T^2,\lambda,(\Lambda,\tvec{S}_T))
 &=& \frac{1}{2} \Bigg[ f_1^{\phantom{\perp}\!\!}
   + S_T^i \epsilon^{ij} k^j \frac{1}{m}\, f_{1T}^\perp
   + \lambda \Lambda\, g_1^{\phantom{\perp\!\!}}
   + \lambda\, S_T^i k^i \frac{1}{m}\, g_{1T}^{\phantom{\perp\!\!}} 
  \Bigg] \, ,
\\
\tilde\rho(x,\tvec{k}_T^2,\tvec{s}_T,\tvec{S}_T)
 &=& \frac{1}{2} \Bigg[ f_1^{\phantom{\perp\!\!}}
   + S^i_T \epsilon^{ij} k^j \frac{1}{m}\, f_{1T}^\perp
   + s^i_T \epsilon^{ij} k^j \frac{1}{m}\, h_{1}^\perp
   + s^i_T S^i_T h_1^{\phantom{\perp\!\!}}
\nonumber \\
  \label{piet-distr2}
 && \hspace{0.7em}
 {}+ s^i_T (2 k^i k^j - \tvec{k}^2 \delta^{ij}) S^j_T 
       \frac{1}{2m^2}\, h_{1T}^\perp
   +  \Lambda\, s^i_T k^i \frac{1}{m}\, h_{1L}^\perp \Bigg] \, ,
\end{eqnarray}
where the distribution functions depend on $x$ and $\tvec{k}_T^2$.
As first outlined in Ref.\cite{Diehl:2005jf} 
and further discussed in a more broad context in 
Ref.\cite{Meissner:2007rx}, the tensor structure 
in (\ref{eq:long}) and (\ref{eq:transv}) are analogs of those in
(\ref{piet-distr1}) and (\ref{piet-distr2}), respectively,
with $\tvec{k}_T$ playing the role of $\tvec{b}$.
However  $\tvec{k}_T$ and $\tvec{b}$ are not conjugate variables, and 
therefore the distributions in the transverse momentum are not Fourier
transform of the impact-parameter dependent distributions.
The analogy between the distributions in the two spaces reads
\begin{eqnarray}
  \label{dictionary}
&& 
f_1^{\phantom{\perp}} \leftrightarrow H , \hspace{9.3em}
f_{1T}^\perp          \leftrightarrow {}- E' , \hspace{6.6em}
g_1^{\phantom{\perp}} \leftrightarrow \tilde{H} ,
\nonumber \\[0.2em]
&& 
h_1^{\phantom{\perp}} \leftrightarrow
    H_T - \Delta_b \tilde{H}_T /(4m^2) \, , \qquad
h_1^\perp      \leftrightarrow {}- (E_T' + 2\tilde{H}_T') \, , \qquad
h_{1T}^\perp   \leftrightarrow 2 \tilde{H}_T'' \, .
\end{eqnarray}
The impact-parameter distributions which would correspond to 
$g_{1T}$ and $h_{1L}^\perp$ are absent because of time-reversal invariance. Therefore the
dipole correlations related to these TMDs are a characteristic feature of intrinsic transverse momentum.
\begin{figure}[t]
\centerline{\hspace{-1 cm}\psfig{file=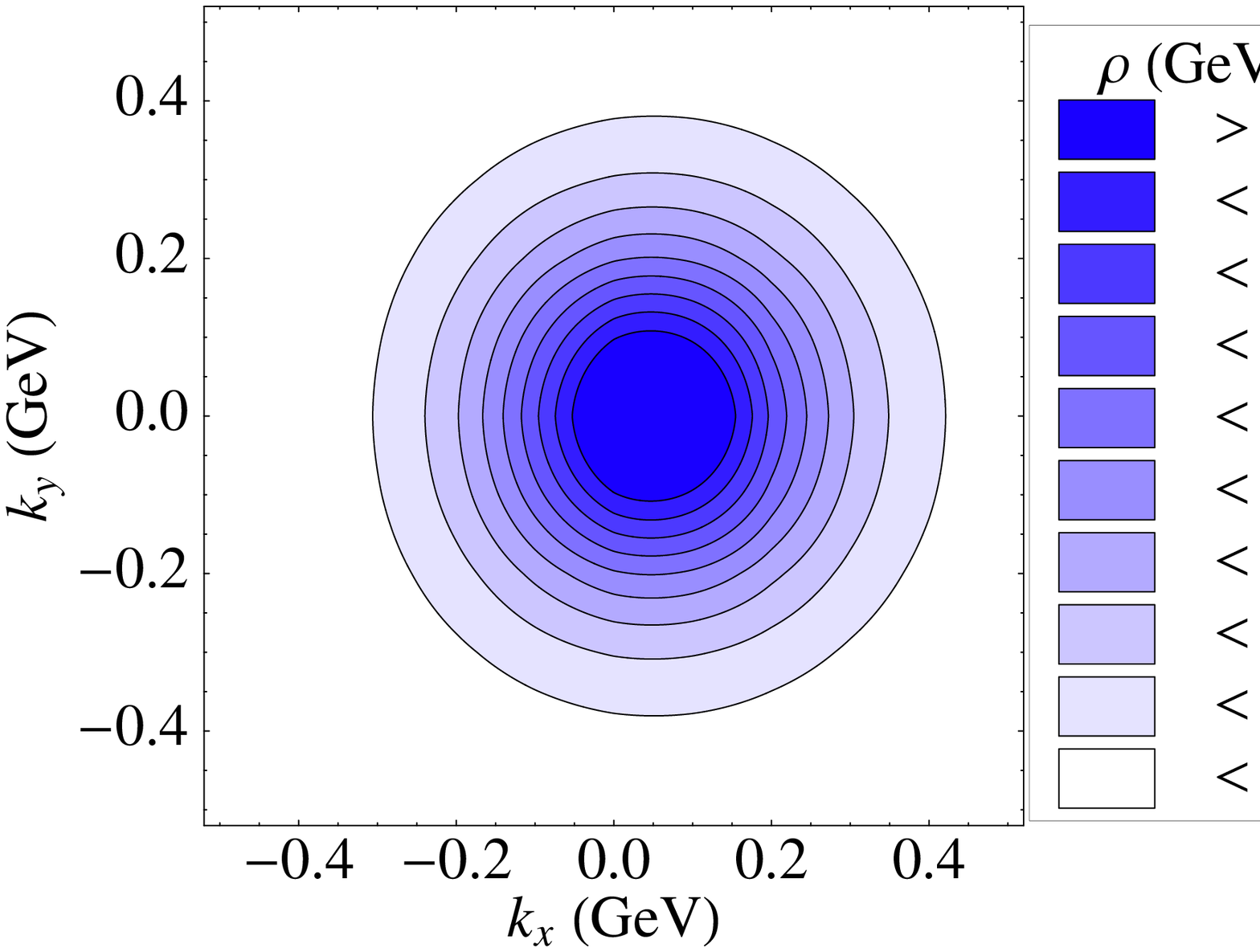, width=4.5 cm}
\hspace{1 cm}
\psfig{file=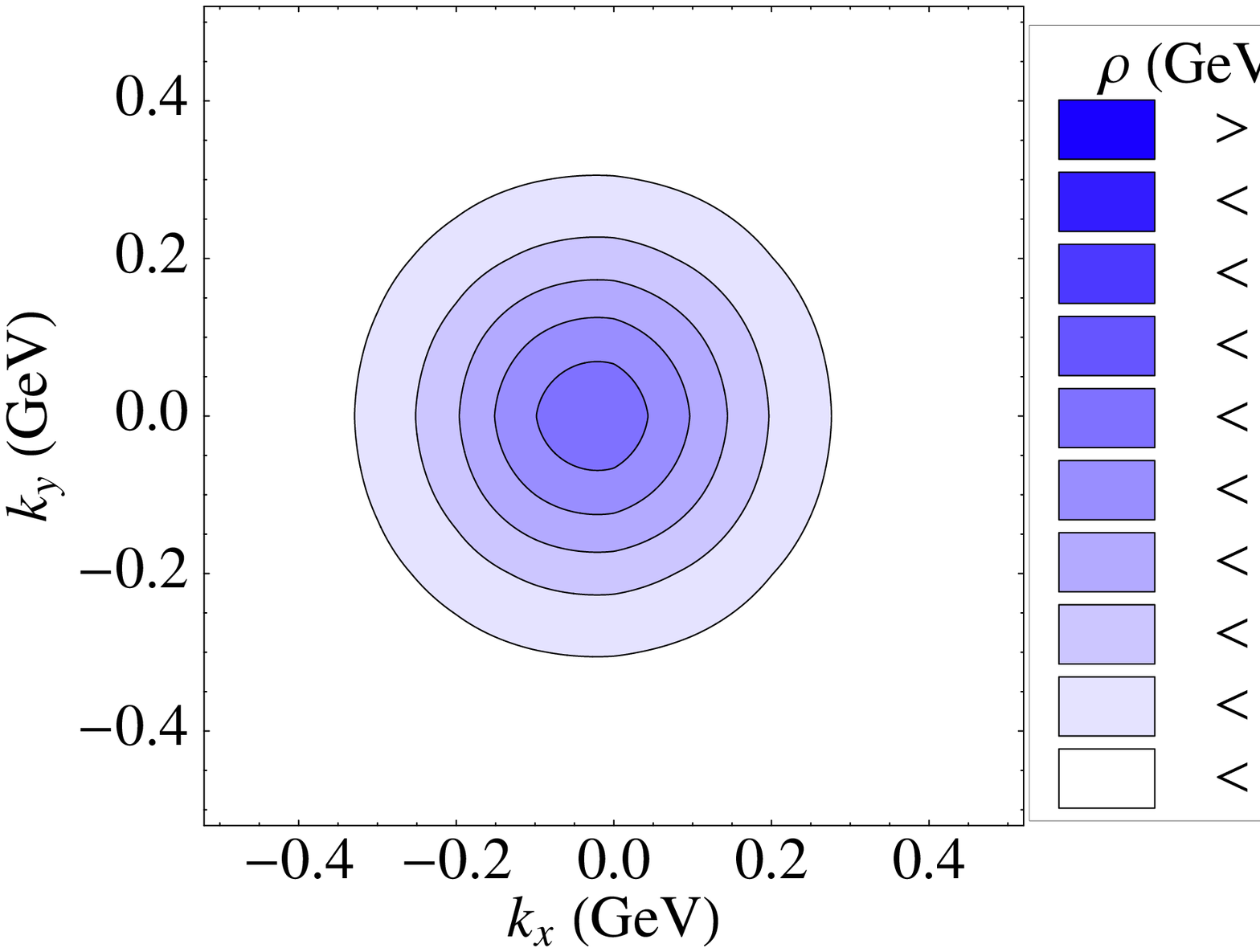, width=4.5 cm}}
\vspace{-12pt}
\caption{Quark densities in the $\tvec{k}_T$ plane
for longitudinally polarized quarks in a transversely polarized 
proton for up (left panel ) and down (right panel) quark.
\protect\label{fig2}}
\end{figure}
The results in the light-cone quark model of Ref.\cite{Pasquini:2008ax} 
for the densities 
with longitudinally polarized quarks in a transversely polarized 
proton are shown in Fig.~\ref{fig2}.
The sideways shift in the positive (negative) $x$ direction for up (down) quark 
due to the dipole term $\propto \lambda\, S^i k^i \frac{1}{m}\, g_{1T}^{\phantom{\perp\!\!}}$ is sizeable, and corresponds to an average deformation
$\langle \tvec{k}_x^u\rangle=55.8 $ MeV, and  
$\langle \tvec{k}_x^d\rangle=-27.9 $ MeV. 
The dipole distortion  $\propto 
\Lambda\, s^i k^i \frac{1}{m}\, h_{1L}^\perp$ in the case of transversely polarized quarks in a longitudinally polarized proton is equal but opposite in sign, 
since in our model $h_{1L}^\perp=-g_{1T}$.
These model results are supported from a recent lattice 
calculation\cite{Hagler:2009mb}
which gives, 
for the density related to $g_{1T}$,
$\langle \tvec{k}_x^u\rangle=67(5) $ MeV, and  
$\langle \tvec{k}_x^d\rangle=-30(5) $ MeV. For the density
related to  $h_{1L}^\perp$,  they also find shifts of similar magnitude but opposite sign:
$\langle \tvec{k}_x^u\rangle=-60(5) $ MeV, and  
$\langle \tvec{k}_x^d\rangle=15(5) $ MeV.

The LCWF overlap representation of the  T-even TMDs in 
Eq.~(\ref{dictionary}) are given in terms of the same 
combinations of light-cone amplitudes  parametrizing the corresponding GPDs 
at $\xi=0$, but taken for different values of the transverse momenta of the 
quarks.
In particular, TMDs are diagonal in the momentum space of the three quarks 
and are unintegrated over the transverse momentum of the active quark.
On the other side, GPDs are integrated over the transverse momenta of 
all the three quarks, but with a finite transverse-momentum transfer 
between the quarks in the initial and final state.
Therefore, the possibility to establish a direct relationship between TMDs and GPDs exists only in the kinematical limits where the differences 
in the momentum dependence of the light-cone amplitudes vanish.
This is trivially the case when both the TMDs and the GPDs reduce to the ordinary quark distribution functions, i.e.  
\begin{eqnarray}
&&\int{\rm d}\tvec{k}_T^2 \,f_1(x,\tvec{k}_T^2)=f_1(x)=H(x,\xi=0,t=0),\nn
\end{eqnarray}
\begin{eqnarray}
&&\int{\rm d}\tvec{k}_T^2 \,g_{1L}(x,\tvec{k}_T^2)=g_1(x)=\tilde H(x,\xi=0,t=0),
\nn
\end{eqnarray}
\begin{eqnarray}
&&\int{\rm d}\tvec{k}_T^2 \,h_{1}(x,\tvec{k}_T^2)=h_1(x)=H_T(x,\xi=0,t=0),
\end{eqnarray}
where $f_1(x)$, $g_1(x)$ and $h_1(x)$ are the unpolarized, helicity and transversity quark distributions, respectively.
\newline
\noindent
Moreover, within the light-cone CQM we find the following non-trivial 
relation
\begin{eqnarray}
\label{eq:tmd_gpd}
\int{\rm d}\tvec{k}_T^2  \,h_{1T}^{\perp}(x,\tvec{k}^2_T)=\frac{2}{(1-x)^2}\tilde H_T(x,0,0).
\end{eqnarray}
A similar relation holds also within the diquark spectator 
model\cite{Meissner:2007rx}, with a factor 3 instead of 2. 
The difference  in the two model calculations supports the conclusions 
of Ref.\cite{Meissner:2007rx} that 
the relationship between $h_{1T}^\perp$ and $\tilde H_T$ cannot be established 
in a model-independent way, even when we restrict ourselves 
to the simplest situation with only valence quark contribution.
\newline
\noindent
Furthermore, in the present light-cone quark model one can write down
relations also among TMDs. Although in QCD the various TMDs are all independent of each other and describe different aspects of the nucleon structure, 
it is quite natural to encounter such relations in simple models limited to
the valence-quark contribution. For a more detailed discussion on this point, 
we refer to\cite{Avakian:2008dz,Schweitzer:proc}.
\begin{figure}[b]
\centerline{\psfig{file=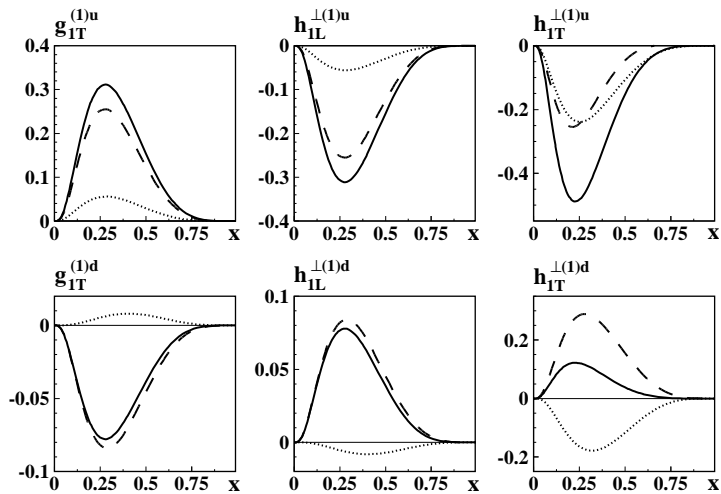, width=10 cm}}
\vspace*{8pt}
\caption{Transverse moments of 
TMDs
as function of $x$ for up (upper panels) and down (lower panels) quark.
In all panels the solid curves show the total results, 
sum of the partial wave contributions.
In the case of $ g_{1T}^{(1)}$ and $h_{1L}^{\perp(1)}$ the dashed and dotted curves give the results from the S-P and P-D interference terms, respectively. 
In the case of $h_{1T}^{\perp(1)}$, the dashed curve  is the result from P-wave interference, and the dotted curve is due to the interference of S and D waves.
\protect\label{fig3}}
\end{figure}
\newline
Finally, in Fig.~\ref{fig3} is shown the interplay between the different 
partial-wave contributions to the transverse moments  $ g_{1T}^{(1)}$, $h_{1L}^{\perp(1)}$ and $h_{1T}^{\perp(1)}$. They are defined as the integrals in $\tvec{k}^2_\perp$ of the TMDs  multiplied by $\tvec{k}^2_\perp/2m^2$.
While the functions $ g_{1T}^{(1)}$ and $h_{1L}^{\perp(1)}$ are dominated by the
 contribution due to P-wave interference, in the case of $h_{1T}^{\perp(1)}$
the contribution from the D wave is amplified through the interference 
with the S wave. The total results for up and down quarks
obey the SU(6) isospin relation, i.e. the functions for up quarks 
are four times larger than for down quark and with opposite sign.
This does not apply to the partial-wave contributions, as it is evident in 
particular for the terms containing D-wave contributions.

\section{Results for azimuthal SSAs}

\begin{figure}[b]
\centerline{
 \hspace{5mm}
\psfig{file=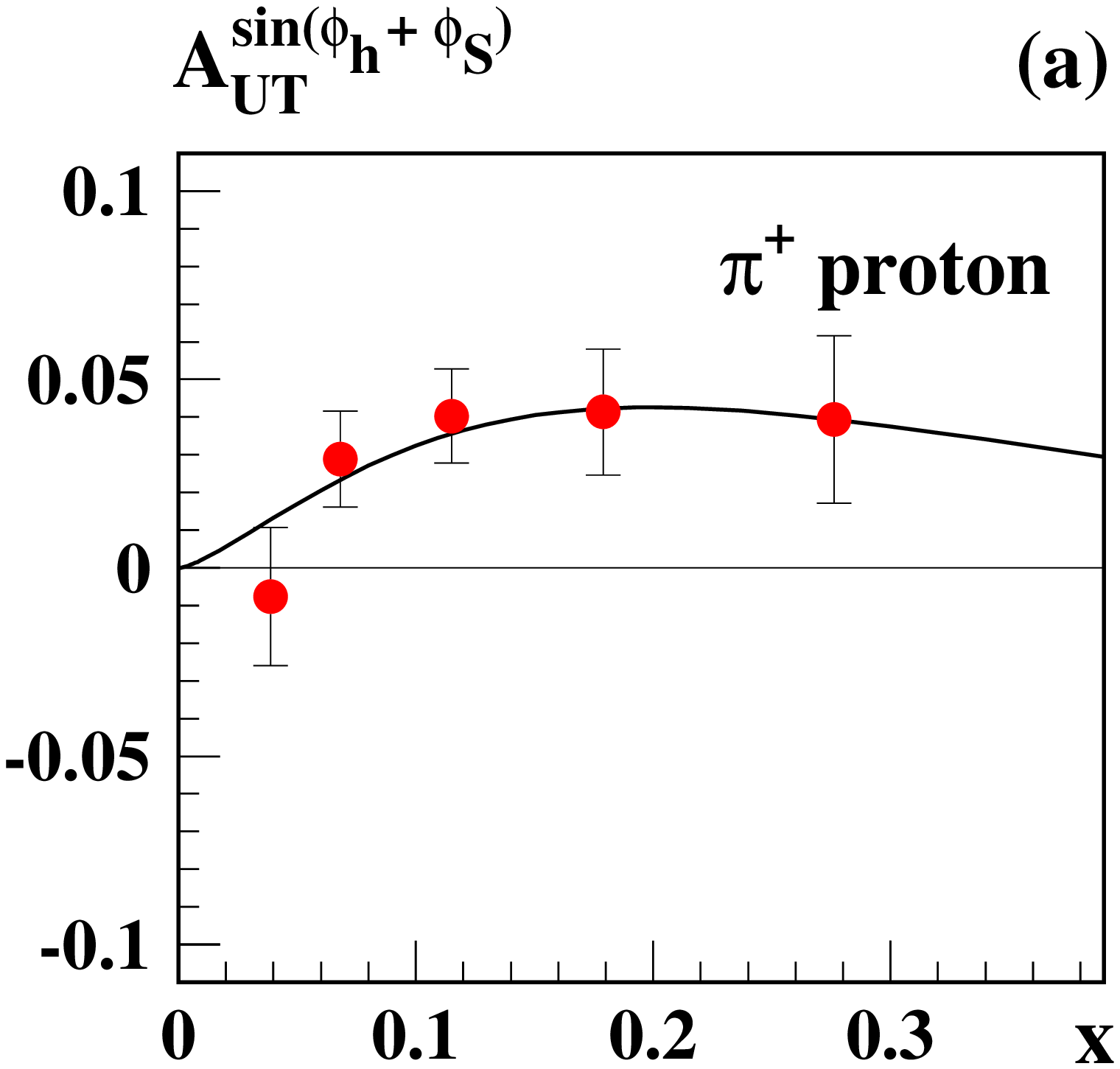, height=4.1 cm}
 \hspace{-11mm}
\psfig{file=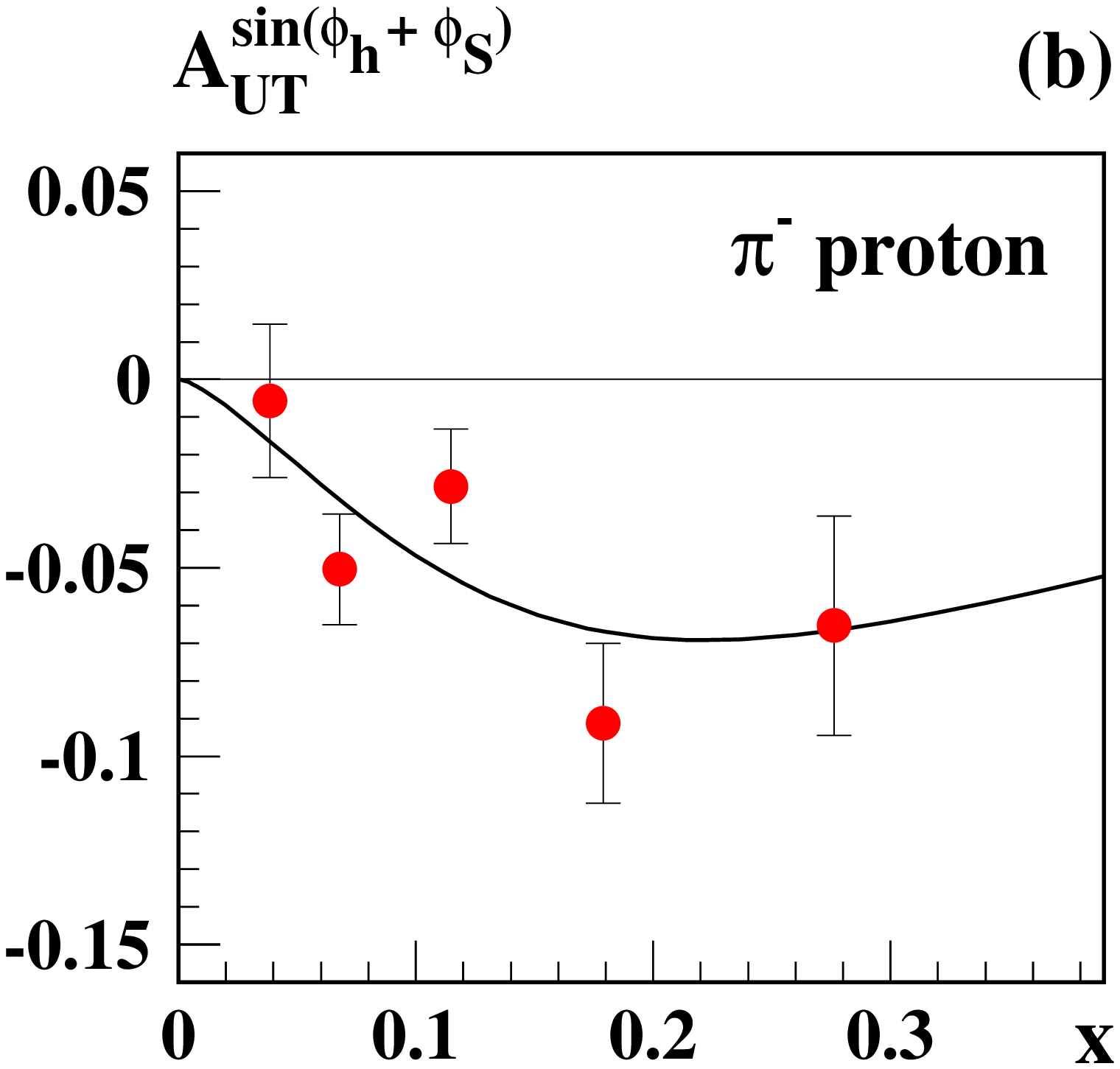, height=4.1 cm}
 \hspace{-12mm}
\psfig{file=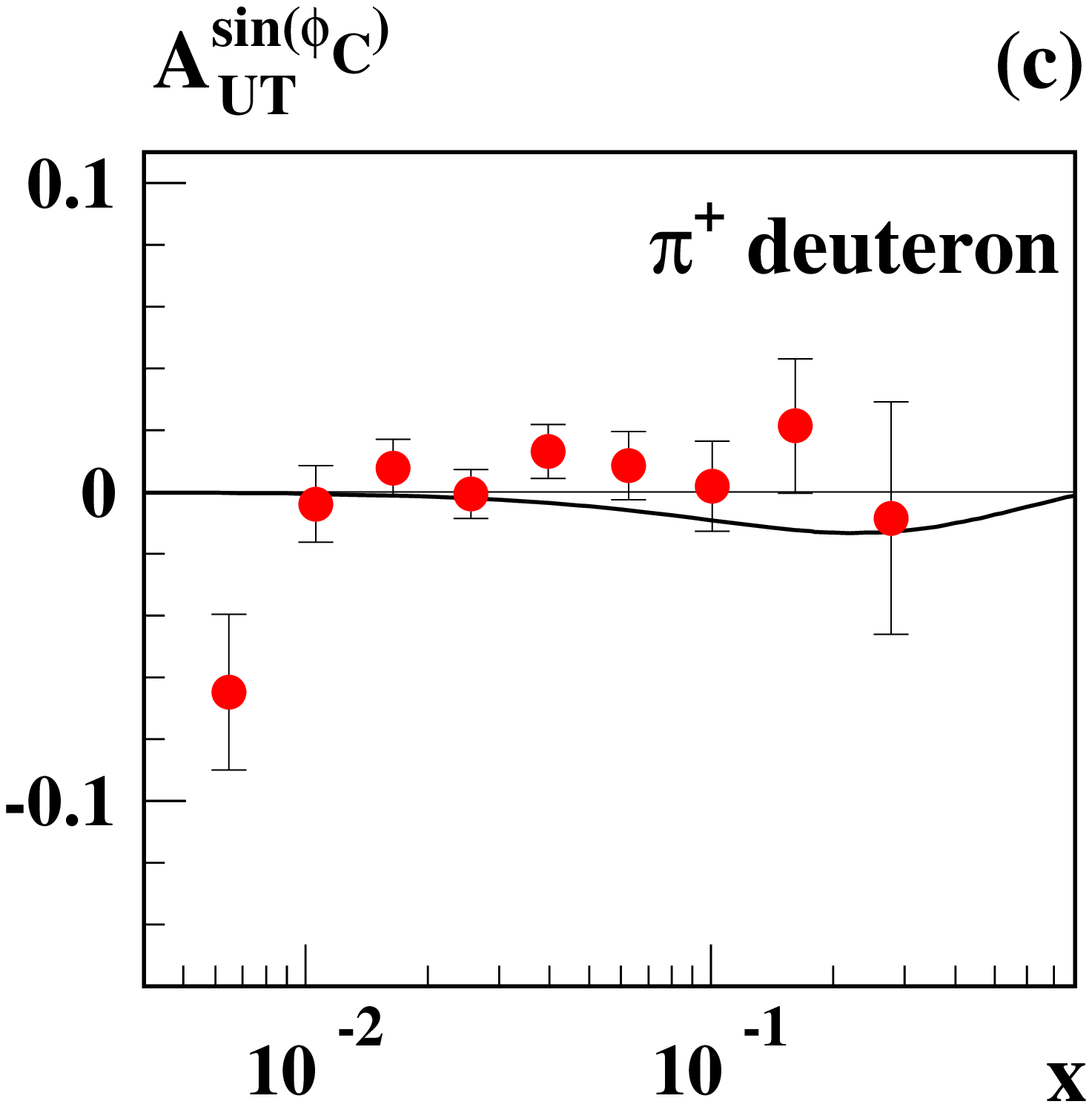, height=4.1 cm}
 \hspace{-11mm}
\psfig{file=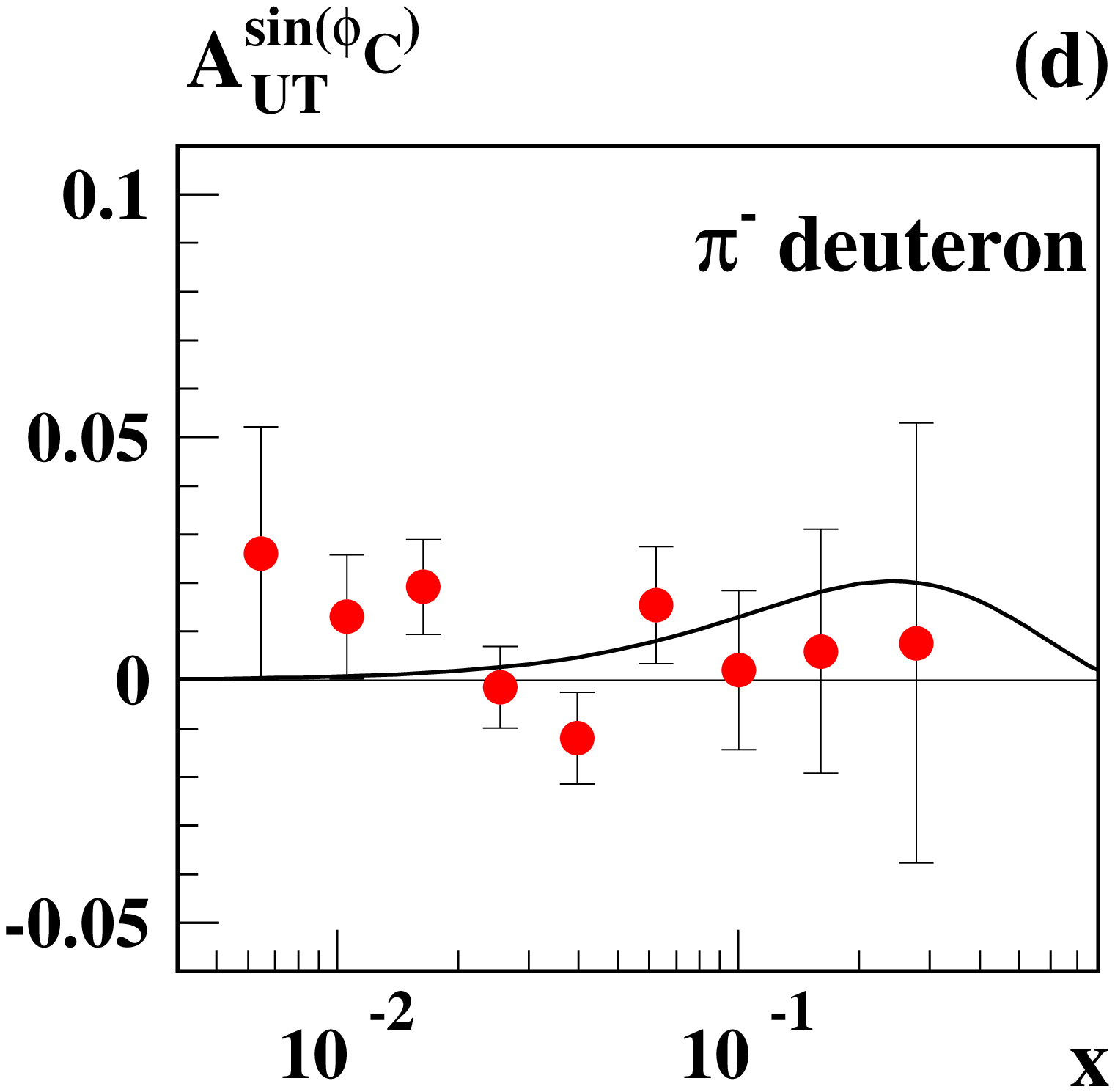, height=4.1 cm}
}
	\caption{\label{fig4}
	The single-spin asymmetry 
        $A_{UT}^{\sin(\phi_h+\phi_S)}\equiv-A_{UT}^{\sin\phi_C}$ in DIS
	production of charged pions 
	off proton and deuterium targets, as function 
        of $x$. The theoretical curves are obtained on the basis of the 
        light-cone CQM predictions for $h_1(x,Q^2)$ from
	Ref.~\protect\cite{Pasquini:2005dk,Pasquini:2008ax}.
        The (preliminary) proton target data are 
        from HERMES \protect\cite{Diefenthaler:2005gx}, 
	the deuterium target data are from COMPASS \protect\cite{Alekseev:2008dn}.}
\end{figure}
\begin{figure}[t]
\vspace{1 truecm}
\centerline{
 \hspace{7mm}
 \includegraphics[height=3.15cm]{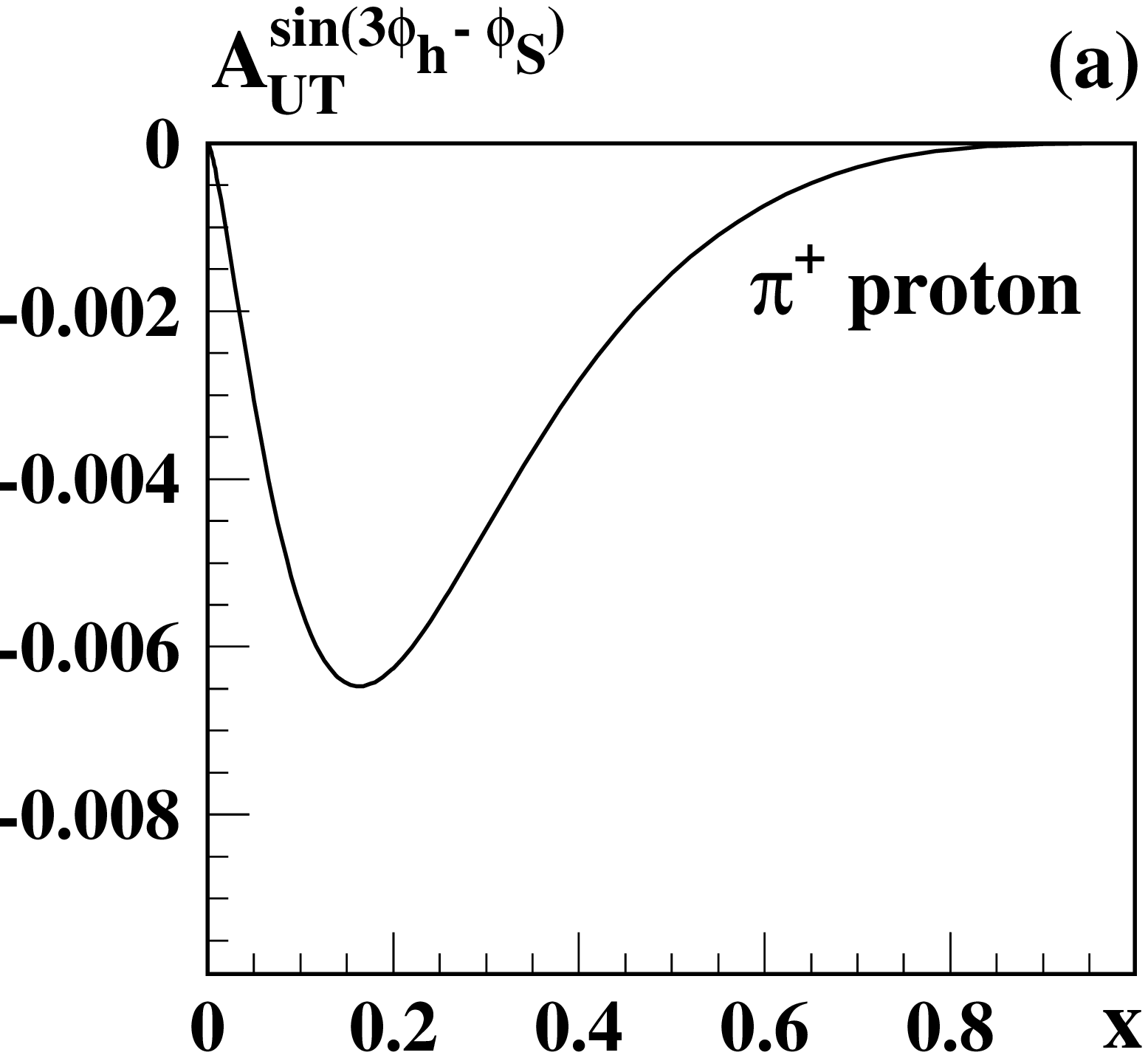}
 \hspace{-11mm}
 \includegraphics[height=3.15cm]{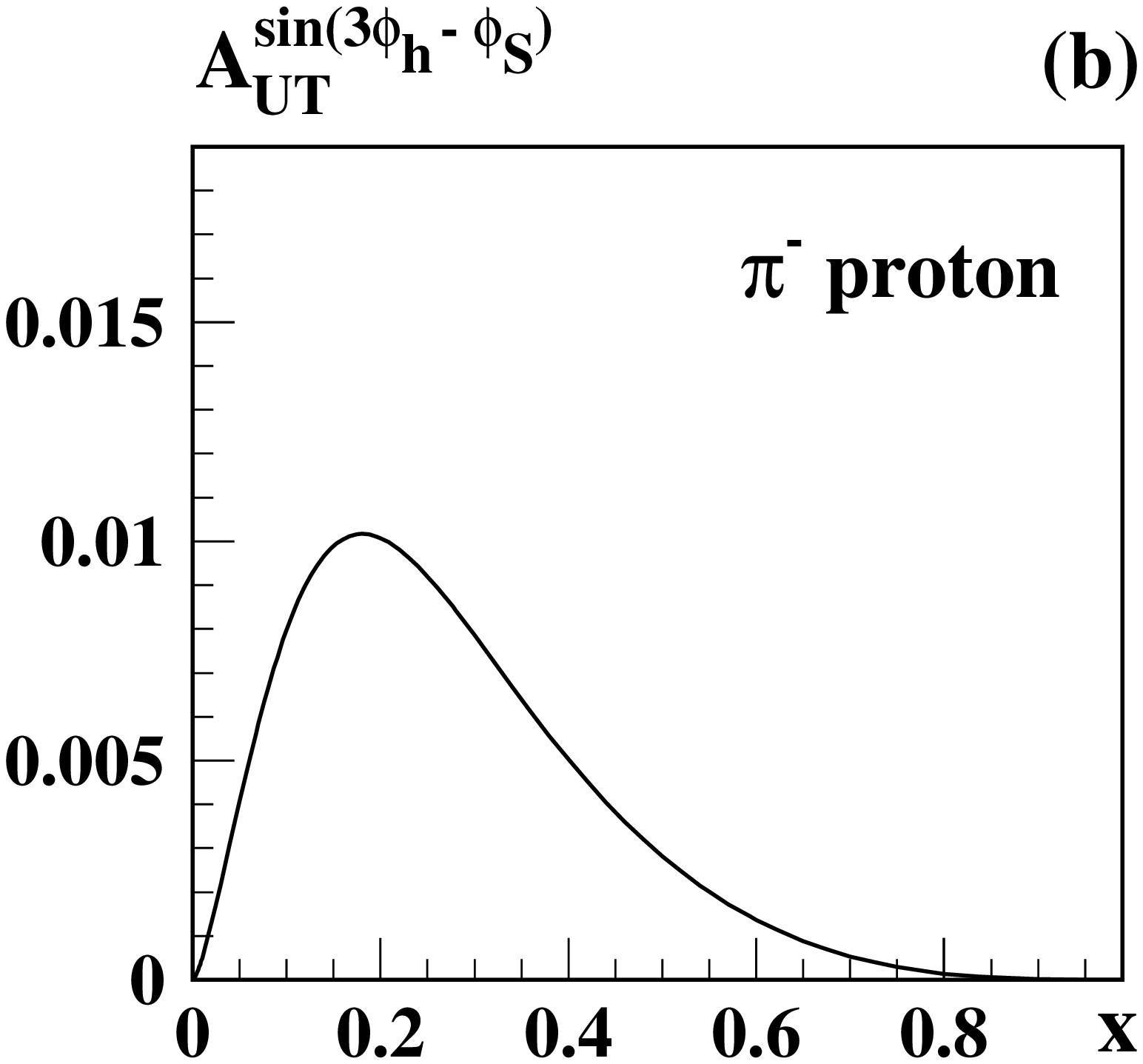}
 \hspace{-10mm}
 \includegraphics[height=3.15cm]{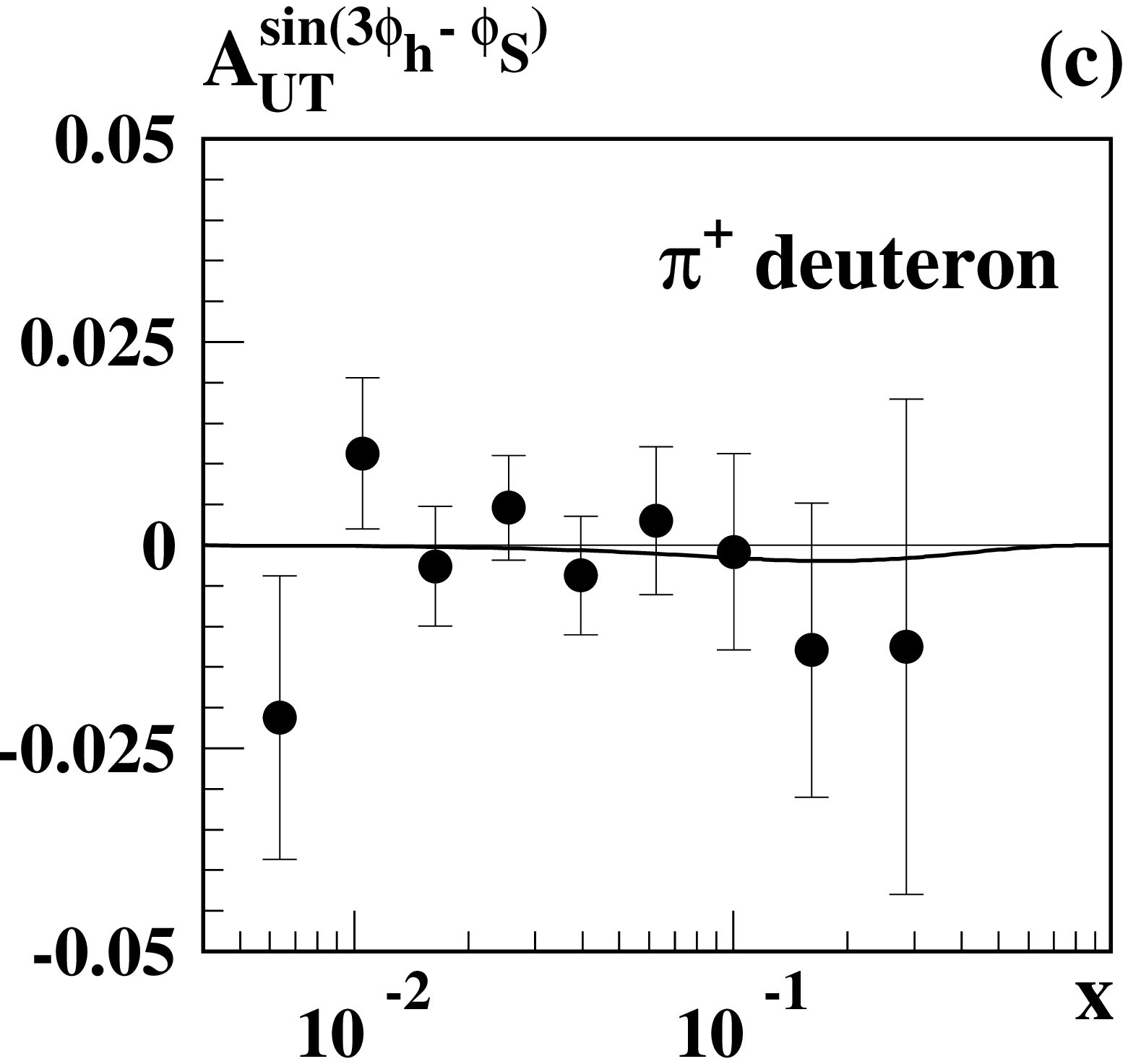}
 \hspace{-10mm}
 \includegraphics[height=3.15cm]{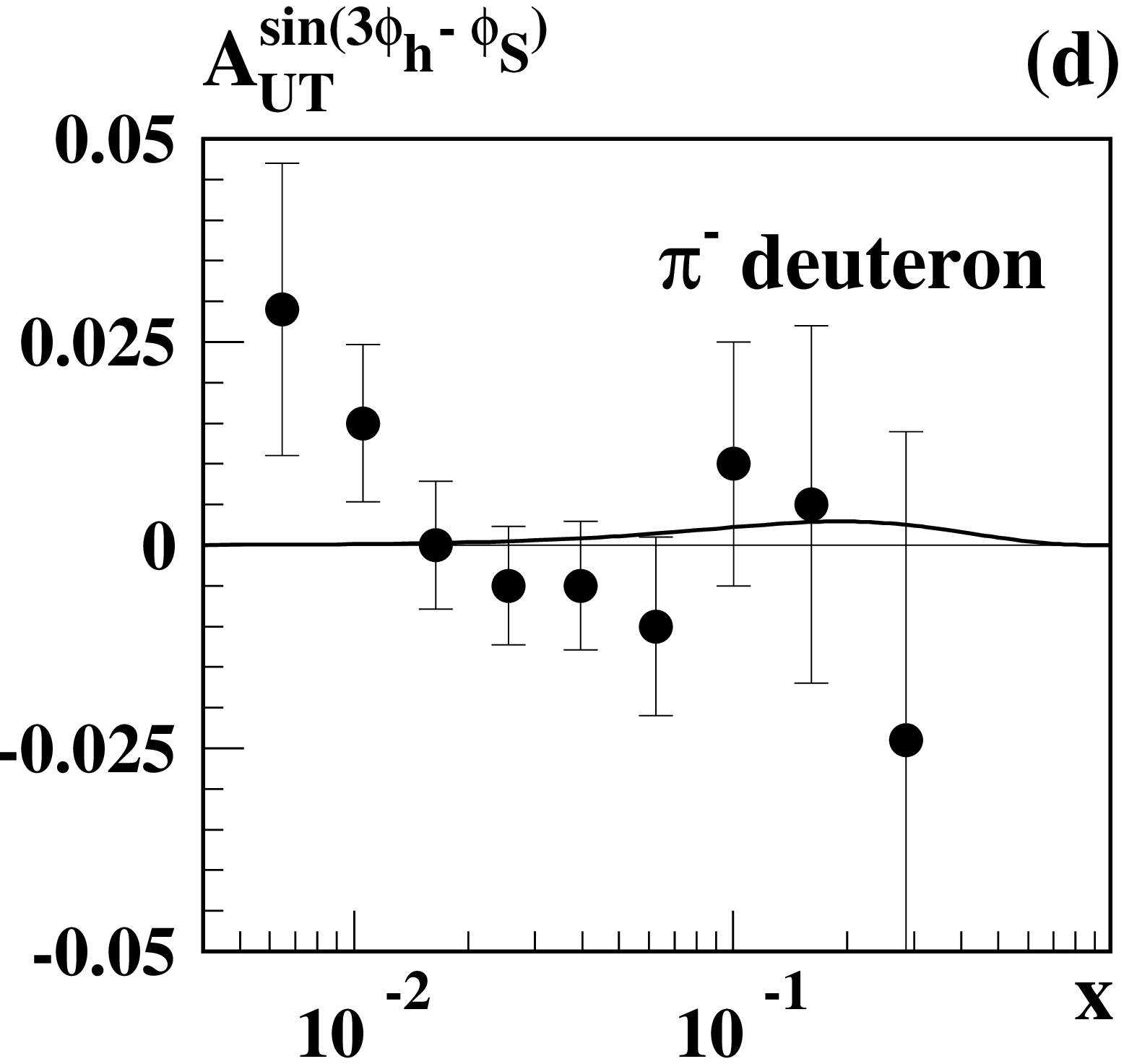}
}
{
\vspace{-0.4 truecm}
    	\caption{\label{autbis}
	The single-spin asymmetry $A_{UT}^{\sin(3\phi_h-\phi_S)}$ in DIS
	production of charged pions off proton and deuterium targets, 
	as function of $x$.
	The theoretical curves are obtained by evolving
	the light-cone CQM predictions for $h_{1T}^{\perp(1)}$ of
	Ref.~\protect\cite{Pasquini:2008ax} to $Q^2=2.5$ GeV$^2$, using
	the $h_1$ evolution pattern.
        The preliminary COMPASS data are from Ref.~\protect\cite{Kotzinian:2007uv}.
}
}
\end{figure}

In Ref.\cite{Boffi:2009sh} the present results for the T-even TMDs 
were applied to estimate azimuthal asymmetries in SIDIS, discussing
the range of applicability of the model, especially with regard to the scale
dependence of the observables and the transverse-momentum dependence of the 
distributions.
Here we review the results  for the Collins asymmetry
 $A_{UT}^{\sin(\phi+\phi_S)}$  and for  $A_{UT}^{\sin(3\phi-\phi_S)}$, due to the Collins fragmentation function and to the chirally-odd TMDs $h_1$,  and 
$h_{1T}^\perp$, respectively.
In both cases, we use the results extracted in \cite{Efremov:2006qm}
for the Collins function.
In the denominator of the asymmetries we take $f_1$ from\cite{Gluck:1998xa}
and the unpolarized fragmentation function from\cite{Kretzer:2000yf}, both valid at the scale $Q^2=2.5$ GeV$^2$.
\newline
In Fig.~\ref{fig4} the results for the Collins asymmetry
in DIS production of charged pions off proton and deuterium targets are shown 
as function of $x$.
The model results for $h_1$ evolved from the low hadronic scale
of the model to $Q^2=2.5 $ GeV$^2$ ideally describe the HERMES 
data\cite{Diefenthaler:2005gx} for a proton target (panels (a) and (b) of Fig.~\ref{fig4}).
This is in line with the favourable comparison 
between our model predictions\cite{Pasquini:2005dk}  and the phenomenological extraction 
of the transversity and the tensor charges in Ref.\cite{Anselmino:2007fs}.
Our results are compatible also with the COMPASS data\cite{Alekseev:2008dn}
for a deuterium target (panels (c) and (d) of Fig.~\ref{fig4}) which extend down to much lower values of $x$.
\newline
In the case of the asymmetry  $A_{UT}^{\sin(3\phi-\phi_S)}$ we face the question
how to evolve $h_{1T}^{\perp(1)}$ from the low scale of the model to the relevant
experimental scale.
Since exact evolution equations are not available in this case, we ``simulate'' the evolution of  $h_{1T}^{\perp(1)}$ by evolving it according to the transversity-evolution pattern. 
Although this is not the correct evolution pattern, 
it may give us a rough insight on the possible size of effects 
due to evolution 
(for a more detailed discussion we refer to\cite{Boffi:2009sh}).
The evolution effects give 
smaller  asymmetries in absolute value and shift the peak at lower $x$ values
in comparison with the results obtained without evolution.
The results shown in Fig. \ref{autbis} are also much smaller than the 
bounds allowed by positivity, 
$|h_{1T}^{\perp(1)}|\leq\frac12\,(f_1(x)-g_1(x)) $, and
constructed using parametrizations of the unpolarized and helicity distributions at $Q^2=2.5$ GeV$^2$.
Measurements in  range $0.1\lesssim x \lesssim 0.6$ are planned with the CLAS 12 GeV upgrade\cite{Avakian-LOI-CLAS12} 
and will be able to discriminate between these scenarios.
There exist also preliminary 
deuterium target data\cite{Kotzinian:2007uv}
which are compatible, within error bars,
with the model predictions both at the hadronic and the evolved scale.

\section*{Acknowledgments}
The authors are  grateful to  the organizers for the invitation to the 
Workshop on ``Recent Advances in Perturbative QCD and Hadronic Physics'', ECT*, Trento (Italy), 
in honor of Professor  A. V. Efremov to whom we also express the best wishes. 
This work is part of the activity HadronPhysics2, Grant Agreement n. 227431, 
under the Seventh Framework Programme of the European Community, and 
is  supported in part by DOE contract DE-AC05-06OR23177.



\bibliographystyle{aipproc}   



\end{document}